\documentclass[12pt,letter]{article}

\usepackage{graphicx, epsfig, color}
\usepackage{amsmath,amssymb,hyperref}

\def\to{\rightarrow}

\def\bi{\begin{itemize}}
\def\ei{\end{itemize}}

\def\sps1ap{SPS1a$^\prime$}
\def\c1p{C1$^\prime$}

\def\tst{\tilde t}

\def\tg{\tilde g}

\def\tw{\widetilde W}
\def\tz{\chi}

\def\be{\begin{equation}}  
\def\ee{\end{equation}}  
\def\bea{\begin{eqnarray}}  
\def\eea{\end{eqnarray}}  
\def\beas{\begin{eqnarray*}}  
\def\eeas{\end{eqnarray*}}

\begin{document}
\begin{titlepage}
\begin{flushright}
CETUP2016-006
\end{flushright}

\vspace{0.5cm}
\begin{center}
{\Large \bf SUSY under siege from \\ 
direct and indirect WIMP detection experiments
}\\ 
\vspace{1.2cm} \renewcommand{\thefootnote}{\fnsymbol{footnote}}
{\large Howard Baer$^1$\footnote[1]{Email: baer@nhn.ou.edu }, Vernon Barger$^2$\footnote[2]{Email: barger@pheno.wisc.edu } 
and Hasan Serce$^1$\footnote[3]{Email: serce@ou.edu }
}\\ 
\vspace{1.2cm} \renewcommand{\thefootnote}{\arabic{footnote}}
{\it 
$^1$Department. of Physics and Astronomy,
University of Oklahoma, Norman, OK 73019, USA \\
}
{\it 
$^2$Department. of Physics,
University of Wisconsin, Madison, WI 53706, USA \\
}

\end{center}

\vspace{0.5cm}
\begin{abstract}
\noindent 
We examine updated prospects for detecting WIMPs in supersymmetric models
via direct and indirect dark matter search experiments. 
We examine several historical and also still viable scenarios: 
projections for well-tempered neutralinos (WTN), 
projections from the MasterCode (MC), BayesFits (BF) 
and Fittino (FO) collaborations,
non-thermal wino dark matter (NThW)
and finally mixed axion-higgsino dark matter from SUSY with 
radiatively-driven naturalness (RNS). 
The WTN is ruled out by recent limits from XENON and LUX collaborations. 
The NThW scenario, previously on tenuous ground due to gamma-line searches, 
appears also ruled out by recent combined Fermi-LAT/MAGIC limits combined with
new HESS results from continuum gamma rays.
Substantial portions of MC parameter space and 1 TeV higgsino parameter space from BF 
group are ruled out. 
The 100-300 GeV higgsino-like WIMP from RNS survives due to its possible
depleted local abundance (where the axion may make up the bulk of dark matter).
Projections from ton-scale noble liquid detectors should discover or rule
out WIMPs from the remaining parameter space of these surviving models. 

\vspace*{0.8cm}

\end{abstract}

\end{titlepage}

\section{Introduction}

Supersymmetric models of particle physics have long generated excitement due to their
ability to tame the naturalness or hierarchy problem associated with quadratic divergences in 
the Higgs mass~\cite{wittenkaul}. 
These models actually receive indirect support from experiment in that 1. 
the measured values of the gauge couplings from LEP unify to a common value at 
$m_{\rm GUT}\simeq 2\times 10^{16}$ GeV under Minimal Supersymmetric Standard Model (MSSM)
renormalization group (RG) evolution~\cite{gauge}, 
2. the measured value of the top quark mass is in the right range to
trigger a radiative breakdown of electroweak symmetry~\cite{rewsb} and
3. the measured value of the Higgs boson mass~\cite{lhc_higgs}
falls squarely within the narrow allowed window required by the MSSM, namely $m_h \lesssim 135$ GeV~\cite{mhiggs}.
In addition, the lightest SUSY particle (LSP) is expected to be absolutely stable under conservation of
$R$-parity which is highly motivated both by theoretical unification issues and also by the need to
stabilize the proton. In this case, then the LSP -- assumed here to be the lightest neutralino of SUSY, $\tz_1$ -- 
presents an excellent candidate for cold dark matter. Simple calculations of its relic abundance indicate
about the right level of thermal dark matter production in the early universe to gain accord with measured values --
a situation known as the WIMP miracle.

Thus, WIMPs (weakly interacting massive particles) from supersymmetric models have long been
an important target for dark matter hunters~\cite{hunt}.
However, lately this long-dominant paradigm appears to be under considerable siege due to:
\begin{itemize}
\item lack of SUSY signals at the CERN Large Hadron Collider (LHC)~\cite{lhc_s} and 
\item the rather high value of $m_h\simeq 125$ GeV requires TeV-scale highly mixed top squarks, 
a situation in conflict with some early evaluations of SUSY electroweak naturalness~\cite{oldnat,Papucci:2011wy} and
\item the lack of any (definitive, verifiable) WIMP signal in either direct or indirect dark matter detection 
experiments~\cite{Baudis:2015mpa}.
\end{itemize}

Given the above conflicting currents, it is incumbent upon theorists to take occasional stock of the 
theory vs. experiment situation with regard to which theoretical models are excluded by data, 
which (if any) are allowed, how plausible the surviving models are, and what remains to be done 
to verify or exclude the surviving models.
In this paper we present such an evaluation. We focus our attention on several recent evaluations of SUSY
model parameter space with regard to direct and indirect dark matter detection. 
These include:
\begin{itemize}
\item models of well-tempered neutralinos (WTN)~\cite{ArkaniHamed:2006mb},
\item the MasterCode (MC) evaluation of Constrained Minimal Supersymmetric Standard Model (CMSSM) 
parameter space~\cite{Buchmueller:2010ai},
\item the BayesFIT (BF) group evaluation of CMSSM parameter space~\cite{Roszkowski:2014wqa},
\item the Fittino (FO) group evaluation of CMSSM parameter space~\cite{Bechtle:2015nua},
\item projections for non-thermal wino-like WIMPs (NThW),
\item projections from SUSY models with radiatively-driven naturalness (RNS) and a higgsino-like WIMP\cite{Baer:2011ec,Baer:2013vpa,Bae:2015jea} and
\item projections from the 19 free weak scale parameter phenomenological MSSM or pMSSM~\cite{Cahill-Rowley:2014boa}.
\end{itemize}
The first five of these models generally assume the (thermally and non-thermally produced) 
relic abundance of SUSY WIMPs saturates the measured dark matter abundance. The fifth model
requires naturalness in both the electroweak and QCD sectors of the theory and thus 
includes two dark matter particles: a higgsino-like WIMP required by electroweak naturalness
and an axion which is required in QCD for a natural solution to the strong CP problem.
The pMSSM evaluations require the thermally-produced WIMP abundance to lie at or below the measured value 
$\Omega_{\tz_1}^{\rm TP}h^2\le 0.12$.

The above SUSY models are confronted by updated experimental exclusion plots. 
These include:
\begin{itemize}
\item updated spin-independent (SI) scattering limits from 447 days of XENON100~\cite{Aprile:2016swn}, PandaX~\cite{Tan:2016zwf} 
and 332 lives days of exposure from the LUX experiment~\cite{Akerib:2016vxi},
\item improved spin-dependent (SD) scattering limits on dark matter annihilations in the Sun from IceCube~\cite{Aartsen:2016exj}, 
\item new combined indirect detection (IDD) limits from Fermi-LAT and MAGIC collaborations on gamma rays arising from
WIMP annihilations into $W^+W^-$ states in dwarf spheroidal galaxies~\cite{Ahnen:2016qkx} and
\item search for WIMP annihilations in the galactic center via ten years of data from the 
HESS collaboration~\cite{::2016jja}.
\end{itemize}
Along with the above excluded regions, it is worthwhile to confront the theoretical expectations
against projections from future direct and indirect detection searches. 
The wide variety of new and upgraded WIMP search experiments are aiming towards ever greater sensitivity 
which promises to either discover SUSY or other WIMP dark matter or else exclude many compelling models.

In accord with our goal of an updated assessment of theory vs. experiment on SUSY WIMP dark matter, in
Sec. \ref{sec:models} we review some of the major features of the above listed SUSY WIMP models. 
In Sec. \ref{sec:SI}, we compare current limits for SI  direct dark matter detection 
against projections from the various models. 
The case of SD WIMP detection is shown in Sec. \ref{sec:SD}.
In Sec. \ref{sec:IDD}, we show results from IDD of WIMPs from searches for excesses in continuum gamma
ray spectra emanating from galactic WIMP-WIMP annihilation.
One useful feature of our results is that projections from the various models can be compared
on a single plot. Furthermore, each model is projected onto each different search plot
so that the strengths of different search techniques can be compared.
In Sec. \ref{sec:conclude} we present a summary and conclusions. 

\section{Some recent models for SUSY WIMP dark matter}
\label{sec:models}

\subsection{Well-tempered neutralinos:}

The well-tempered neutralino (WTN) is a neutralino where the relative bino-, wino- and higgsino-
components are adjusted to just the right values such that the calculated thermally-produced (TP) 
neutralino abundance $\Omega_{\tz_1}^{\rm TP}h^2$ matches the measured value. 
While proposed on general grounds in Ref.~\cite{ArkaniHamed:2006mb}, the WTN arose earlier in the context
of the hyperbolic branch/focus point region of the CMSSM. The hyperbolic branch~\cite{Chan:1997bi} is the contour
of fixed, small $\mu$ values in $m_0$ vs. $m_{1/2}$ space of the CMSSM model where 
$m_0$ is taken to be such a large value that $m_{H_u}^2$ barely runs to negative
values at the weak scale so that electroweak symmetry is barely broken. 
The focus point (FP)~\cite{Feng:1999mn,Feng:1999zg}
consists of flat contours of constant $\mu$ values -- for fixed $m_{1/2}$ --
where for a large range of $m_0$ values, the value of $m_{H_u}^2$ is run to
the same weak scale values ($m_{H_u}^2$ is focused in its running to the same focal point for
a large range of $m_0$ values). It was emphasized in Ref.~\cite{Feng:1999mn} that this allows for
natural values of TeV-scale squarks and sleptons since the weak scale value of $m_{H_u}^2$
was rather insensitive to the GUT scale value of $m_0$. By dialing $m_0$ to its 
maximal value, $m_{H_u}^2$ becomes somewhat de-focused, but the parameter values do
reach the hyperbolic branch. Since the value of $\mu$ is dialed/tuned to gain the correct
value of $m_Z$, then $\mu$ is found to be small in this HB/FP region just left of
the region where EW symmetry does not break~\cite{Feng:2000gh,Baer:2003jb}.

In the HB/FP region, $m_0$ can be adjusted to its nearly maximal value allowed by 
REWSB such that $\mu$ becomes small and the neutralino becomes well-tempered: of mixed
bino-higgsino variety such that $\Omega_{\tz_1}^{\rm TP}h^2\simeq 0.12$. Since the WTN 
has substantial gaugino and higgsino components, it tends to have a large SI direct 
detection rate since the $\tz_1-\tz_1-h$ coupling depends on a product of
gaugino times higgsino components. Also, indirect detection rates tend to be large~\cite{Feng:2000zu,Baer:2004qq} since
the higgsino-like $\tz_1$ has a large thermally averaged self-annihilation cross section 
times velocity $\langle \sigma v\rangle $ into vector boson pairs.
Since the HB/FP tends to occur in the CMSSM for low values of $A_0$, then it tends to 
produce too low a value of $m_h$. If $A_0$ is increased, then the downward $m_{H_u}^2$ RG running
is enhanced and it tends to run to large instead of small negative values. 
This must be compensated for by realizing the HB/FP region at much higher $m_0$
values $\sim 10-30$ TeV~\cite{Baer:2012mv} depending on $m_{1/2}$, and upon which code is used
to calculate $m_h$.
A large variety of SUSY models with universal or non-universal soft terms give rise to
WTNs~\cite{Baer:2006te,Baer:2008ih}. 

\subsection{Mastercode collaboration:}

The MasterCode (MC) collaboration~\cite{Buchmueller:2010ai} has assembled a variety of computer codes-- 
SoftSUSY/SSARD for spectra, FeynHiggs, MicroMegas, SUFla and SuperIso-- 
with a goal to calculate a long array of observables in supersymmetric models from
which they calculate a $\chi^2$ value\footnote{Observables include: $\Omega_{\tz_1}^{\rm TP}h^2$, 
$\sigma^{SI} (\tz_1, p)$, $m_h$, $BF(B_{d,s}\to\mu^+\mu^- )$, $BF(b\to s\gamma )$, 
$m_W$ along with $BF(B\to\tau\nu )$, $\epsilon_K$, $R_\ell$, $A_{fb}(b)$, $A_\ell(SLD)$,
$\sigma_{had}^0$ and Atlas/CMS sparticle mass bounds.}. 
The MultiNest code is used to scan around the parameter space. 
Supersymmetric models scanned over include CMSSM, NUHM1, NUHM2 and pMSSM10. 
The best fit regions~\cite{Bagnaschi:2015eha} tend to be dominated by the requirements 
1. to get $\Omega_{\tz_1}^{\rm TP}h^2$ near its measured value, 
2. to obtain $m_h$ within its measured range and 
3. to obtain $a_\mu$ as close as possible to its measured value. 
Requirement \#1 selects out special regions of parameter space 
needed to obtain the measured dark matter relic density: stau, stop or electroweakino 
co-annihilation, well-tempered (mixed bino-higgsino) regions and 
$A/H$ or $h/Z$ resonance annihilation. The pull from $a_\mu$ is towards lighter spectra 
which include light smuons and mu-sneutrinos: this means low values of $m_0$ and $m_{1/2}$ in 
CMSSM parameter space. 
The rather large value of $m_h$ pulls towards non-zero $A_0$ terms and higher $m_0$ and $m_{1/2}$ values.

\subsection{BayesFits collaboration:}

The BayesFits group~\cite{Roszkowski:2014wqa} has assembled a calculational scheme similar to the MC Collaboration, making
use of SoftSUSY interfaced to FeynHiggs, MicroMegas and SuperISO to also examine 
a wide array of observables expected from supersymmetric models. Key observables include: 
the Higgs mass $m_h$, the thermally-produced neutralino relic density $\Omega_{\tz_1}^{\rm TP}h^2$ and
various $B$ decay branching ratios while respecting LHC and SI direct detection bounds.
A key difference is that BayesFits group calculates a Bayesian prior probability density
to evaluate favorable regions of model parameter space. 
They focus on results especially from the CMSSM model but also from NUHM1.

The BF group finds recently that the stau co-annihilation region is only weakly favored
at $2\sigma$ level. More highly favored is the $A/H$ resonance annihilation 
region which tends to occur at large $\tan\beta$ in the CMSSM where $m_A\sim 2 m_{\tz_1}$
and the $A/H$ decay width is enhanced by the large $b$ and $\tau$ Yukawa couplings.
While the BF stau-coannihilation region should be accessible to LHC14 searches with up 
to 300 fb$^{-1}$ of integrated luminosity, the $A/H$ funnel region occurs at 
$m_{1/2}\sim 1.5-2$ TeV. For comparison, $m_{\tg}\sim 2.5 m_{1/2}$ so this corresponds
to $m_{\tg}\sim 4-5$ TeV, well beyond LHC reach. Nonetheless, this region 
is expected ultimately to exhibit discrepancies with the SM value of 
$BF(B_{s,d}\to \mu^+\mu^-)$ where SUSY contributions to the decay mode are enhanced by
large $\tan\beta$ and low $m_A$. A third region is favored at $1\sigma$ level with
$m_{1/2}\sim 2-3.5$ TeV and $m_0\sim 5-10$ TeV where $m_{\tg}\sim 5-8$ TeV. 
This region contains a higgsino-like LSP of mass $\sim 1$ TeV and is essentially the
large $m_0$ remnants of the HB/FP region with $\mu<M_1<M_2<M_3$. 
The 1 TeV higgsino-like LSP should ultimately be detected by ton-scale nobel liquid
direct detection experiments.

\subsection{Fittino collaboration:} 

The Fittino collaboration~\cite{Bechtle:2015nua} has also performed detailed fits to the CMSSM model, 
this time including as well vacuum stability constraints.
They use SPheno/FeynHiggs for the SUSY/Higgs spectrum calculation and 
a Markov chain Monte Carlo search over parameter space using 
Fittino to determine goodness of fit as a $p$-value. 
Constraints from LHC8 with 20 fb$^{-1}$ of data are imposed.
Overall, they conclude that CMSSM is excluded at 90\% CL. 
Nonetheless, the remaining best fit regions are focus point/WTN which merges to 1 TeV higgsino-like 
WIMP at high WIMP mass along with stau co-annihilation and $A$ resonance annihilation.

\subsection{Non-thermal winos:}

The possibility of wino dark matter became exciting with the advent of SUSY breaking models
based on anomaly-mediation~\cite{Randall:1998uk} (AMSB) where a hierarchy of $M_2<M_1<M_3\sim \mu$ is expected. 
The wino-like WIMP $\tz_1$ is close in mass to its charged wino counterparts $\tw_1^\pm$ 
so that both annihilation and co-annihilation combine to produce a predicted thermal
wino abundance that is typically a few orders of magnitude below measured values. 
Soon after the advent of AMSB models, Moroi and Randall~\cite{Moroi:1999zb} proposed {\it non -thermal}
production of wino dark matter via  weak scale moduli field decay in the early universe.
Such non-thermal processes could bolster the thermally produced wino abundance and bring it into
accord with measured values. In addition, it has been proposed that the relic 
wino abundance could be enhanced by~\cite{Baer:2010kd} 2. gravitino production and decay or 3. axino/saxion
production and decay. In the latter case, the wino abundance would be accompanied by
an axion abundance so both WIMPs and axions would be present~\cite{Bae:2015rra}. 
In that case, the winos need not saturate the entire relic abundance.

Relic wino-like WIMPs should annihilate at large rates one-with-another so as to produce large
indirect detection signals. 
In the case where winos do saturate the measured relic abundance, then they are subject to 
strong constraints arising from measured rates for both line and continuum 
gamma ray production from the galactic center and from nearby dwarf galaxies. 
In fact, there are recent claims that such constraints rule out
the possibility of wino-like WIMPs~\cite{Cohen:2013ama,Fan:2013faa}.

\subsection{Higgsino-like WIMPs from radiatively-driven natural SUSY:}

Currently the LHC13 with $\sim 20$ fb$^{-1}$ of integrated luminosity {\it excludes} $m_{\tg}\lesssim 1.9$ TeV within the
framework of various simplified models~\cite{ATLAS:2016kts}. 
This is to be compared with early estimates of upper bounds on sparticle
masses from naturalness~\cite{oldnat} 
which claim $m_{\tg}\lesssim 350$ GeV for fine-tuning parameter $\Delta_{\rm BG}<30$. 
The validity of these upper bounds has been challenged in that they were derived within the context of multi-parameter 
effective theories whereas in more fundamental theories the soft terms are related~\cite{Baer:2014ica} ({\it e.g.} in gravity-mediation, 
the soft terms are all calculable as multiples of $m_{3/2}$ and thus {\it not independent of each other})~\cite{sw}. 

In addition, LHC13 requires $m_{\tst_1}\gtrsim 850$ GeV~\cite{ATLAS:2016jaa} whilst some claims for naturalness required {\it three} third generation squarks
lighter than 500 GeV~\cite{Papucci:2011wy}. The 500 GeV upper bounds have been challenged in that various {\it dependent} 
contributions to the RGEs have been simplified to zero whereas upon inclusion, these terms lead to radiatively-driven naturalness:
for large enough high scale values of up-Higgs soft term, then  $m_{H_u}^2$ is driven radiatively to natural values $\sim -m_Z^2$
at the weak scale~\cite{Baer:2013gva}.

A more model-independent measure of naturalness $\Delta_{\rm EW}$ has been advocated in Ref.~\cite{Baer:2012up,Baer:2012cf}: 
SUSY is electroweak natural if there are no large cancellations on the right-hand-side of the 
weak scale scalar potential minimization condition:
\be 
\frac{m_Z^2}{2} = \frac{m_{H_d}^2 + \Sigma_d^d(j) -(m_{H_u}^2+\Sigma_u^u(k))\tan^2\beta}{\tan^2\beta -1} -\mu^2 
\simeq  -m_{H_u}^2-\Sigma_u^u(k)-\mu^2 
\label{eq:mzs}
\ee
Here, $m_{H_u}^2$ and $m_{H_d}^2$ are squared soft SUSY breaking
Lagrangian terms, $\mu$ is the superpotential higgsino mass parameter,
$\tan\beta =v_u/v_d$ is the ratio of Higgs field
vacuum-expectation-values and the $\Sigma_u^u(k)$ and $\Sigma_d^d(j)$
contain an assortment of radiative corrections, the largest of which
typically arise from the top squarks.  Expressions for the $\Sigma_u^u$
and $\Sigma_d^d$ are given in the Appendix of Ref.~\cite{Baer:2012cf}. 
The fine-tuning measure $\Delta_{\rm EW}$ compares the largest independent contribution on the
right-hand-side (RHS) of Eq.~(\ref{eq:mzs}) to the left-hand-side $m_Z^2/2$.  If
the RHS terms in Eq.~(\ref{eq:mzs}) are individually
comparable to $m_Z^2/2$, then no
unnatural fine-tunings are required to generate $m_Z=91.2$ GeV. 
The main requirements for low fine-tuning ($\Delta_{\rm EW}\lesssim 30$) are then the following\footnote{
The onset of fine-tuning for $\Delta_{\rm EW}\gtrsim 30$ is visually displayed in Ref.~\cite{Baer:2015rja}.}.
\bi
\item $|\mu |\sim 100-300$  GeV~\cite{Papucci:2011wy,Baer:2011ec,Chan:1997bi,Kitano:2006gv,Barbieri:2009ev}
(where $\mu \gtrsim 100$ GeV is required to accommodate LEP2 limits 
from chargino pair production searches).
\item $m_{H_u}^2$ is driven radiatively to small, and not large,
negative values at the weak scale~\cite{Baer:2012up,Baer:2012cf}.
\item The top squark contributions to the radiative corrections
$\Sigma_u^u(\tst_{1,2})$ are minimized for TeV-scale highly mixed top
squarks~\cite{Baer:2012up}.  This latter condition also lifts the Higgs mass to
$m_h\sim 125$ GeV.  For $\Delta_{\rm EW}\lesssim 30$, the lighter top
squarks are bounded by $m_{\tst_1}\lesssim 3$ TeV.
\item The gluino mass which feeds into the $\Sigma_u^u(\tst_{1,2})$ via
RG contributions to the stop masses
is required to be $m_{\tg}\lesssim 4$ TeV, possibly beyond the reach of LHC.
\ei 
SUSY models with these properties have been dubbed radiatively-driven
natural SUSY (RNS) and enjoy low values of $\Delta_{\rm EW}\sim 10-30$. 
In contrast, the presence of a high value of fine-tuning
generally indicates some pathology or missing element within a physical theory.

In RNS SUSY, the LSP is a mainly higgsino-like LSP with mass $m_{\tz_1}\lesssim 300$ GeV (the closer to $m_Z$ the better) 
but with a non-negligible gaugino contribution. They are thermally under-produced. 
Requiring naturalness also in the QCD sector, 
a Peccei-Quinn sector is included so the dark matter consists of an axion-WIMP
admixture (two dark matter particles). WIMPs can be produced both thermally and non-thermally via
axino, saxion and gravitino production and decay in the early universe~\cite{az1} while axions can be
produced via coherent oscillations (production mechanism for the axion dark matter), thermally or via saxion decay 
(in which case they contribute to dark radiation).
The SUSY DFSZ axion has some preference over KSVZ in that it allows for a solution of  the SUSY $\mu$ problem~\cite{kn} 
and can radiatively generate a Little Hierarchy $\mu\ll m_{3/2}$~\cite{radpq}. The complete relic density calculation requires
simultaneous solution of eight coupled Boltzmann equations~\cite{az1}. 
WIMP direct detection rates must all be scaled down~\cite{bottino}
by a factor 
\be
\xi\equiv \Omega_{\tz_1}h^2/0.12
\label{eq:xi}
\ee
due to the fact that the WIMPs comprise only {\it a portion} 
of the local dark matter abundance- the remainder being composed of axions. 
Indirect detection rates are further suppressed since they must be re-scaled by a factor $\xi^2$.

\subsection{pMSSM}
\label{ssec:pmssm}

We will also compare these predictions, at least in the case of SI DD, with projections from the 
19 free weak scale parameter phenomenological MSSM~\cite{Cahill-Rowley:2014boa}. In this model, the authors
advocate predictions which are unprejudiced by renormalization group running from some higher mass scale.
The scans over parameter space typically range up to weak scale soft terms of 4 TeV and are subject to a variety of constraints
including LHC sparticle search limits and that $\Omega_{\tz_1}^{\rm TP}h^2\le 0.12$.
For general projections from a three parameter model involving just electroweak-inos, see Ref.~\cite{Bramante:2015una}.
\section{Spin-independent direct detection}
\label{sec:SI}
\begin{figure}[tbp]
\includegraphics[height=0.474\textheight]{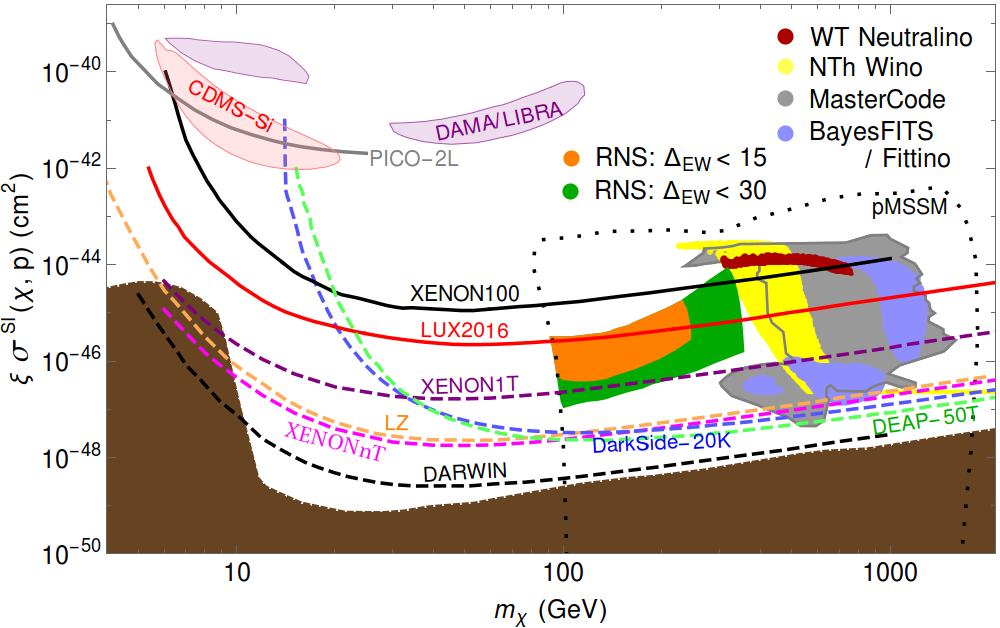}
\caption{Plot of rescaled spin-independent WIMP detection rate $\xi \sigma^{SI}(\chi, p)$ 
versus $m_{\chi}$ from several published results versus current and future reach (dashed) of 
direct WIMP detection experiments. $\xi =1$ ({\it i.e.} it is assumed WIMPs comprise 
the totality of DM) for the experimental projections and for all models {\it except} RNS and pMSSM. 
\label{fig:SI}}
\end{figure}
We first examine a grand overview of prospects for spin-independent SUSY WIMP direct detection. 
In this case, the neutralino-nucleon scattering cross section is dominated by Higgs and squark
exchange diagrams. (Here, most results do not include extensive QCD corrections so 
theory predictions should be accepted to within a factor two unless otherwise noted~\cite{Hisano:2015rsa}. 
Since squark mass limits are now rather high from LHC searches, the Higgs exchange
$h$ diagram usually dominates the scattering amplitude. The results are presented in Fig.~\ref{fig:SI}
in the $\xi\sigma^{SI}(\chi, p)$ vs. $m_{\chi}$ plane. We leave the factor $\xi$ in the $y$-axis
to account for a possible depleted local abundance of WIMPs. 
For the experimental projections and for all models {\it except} RNS and pMSSM, 
it is assumed that $\xi =1$ ({\it i.e.} it is assumed that WIMPs comprise 
the totality of DM).

The lower brown-shaded region denotes the solar neutrino floor: within this region, WIMP
signals would have to contend with a formidable $\nu p$ scattering background. 
In the upper-left, we also show the locus of two anomalous signal regions: from DAMA/LIBRA and from
CDMS-Si. These regions naively appear in conflict with recent limits from XENON and LUX
experiments. For experimental limits, we show the new XENON100 447 live day 
bound~\cite{Aprile:2016swn} (black solid), and the recent 
LUX2016 bound~\cite{Akerib:2016vxi} (which barely supercedes recent PandaX limits~\cite{Tan:2016zwf}). 
In the upper-left, the recent Pico-2L bound is shown~\cite{Amole:2016pye}.
The dashed lines all show projected future reaches of: 
XENON1T~\cite{Aprile:2015uzo}, LZ (with 1 keV cutoff)~\cite{Akerib:2015cja}, XENONnT~\cite{Aprile:2015uzo}, DarkSide-20K~\cite{Agnes:2015lwe}
DEAP-50T~\cite{Amaudruz:2014nsa} and DARWIN noble liquid experiments~\cite{Aalbers:2016jon}.
These latter projections approach to within an order of magnitude of the solar neutrino floor.

For theory models, the maroon-shaded region shows the expected rates for WTNs as derived
from {\it our scan} of the HB/FP region of the CMSSM/mSUGRA model. 
The lower limit arises due to the requirement
of $m_{\tg}>1.9$ TeV in accord with recent LHC13 searches which implies a bino mass $M_1\gtrsim m_{\tg}/7\sim 250$ GeV.
The upper limit arises from requiring a bino-higgsino mixing of at least 10$\%$. For higher $m_{\chi_1}$
values, the LSP becomes more purely higgsino and is no longer tempered, but becomes the 1 TeV higgsino LSP.
The WTN cross sections form a well-known asymptote at $\sigma^{SI}\sim 10^{-44}$ cm$^2$~\cite{Baer:2006te}.
As can be seen, this entire class of models has been ruled out by recent XENON100, PandaX and LUX searches.
The gray-shaded region shows the expected SI-direct detection rates derived by the MC collaboration
(and adapted here from their plots) 
while the blue-shaded regions show expectations from the BayesFits group (also adapted from their plots). 
These projections overlap since we present both groups expectations for the case of the CMSSM model. 
The Fittino preferred regions largely overlap with the results from MC and BF;
for clarity, we do not show these regions.
The lower-left blue/gray
bulge denotes the stau co-annihilation region for $m_{\tz_1}\sim 300-600$ GeV. 
It should  be accessible to LHC14 searches with 300-3000 fb$^{-1}$ and
can also be probed by LZ, XENONnT and DarkSide-20K although perhaps not by XENON1T. 
The lower blue/gray bulge with $m_{\tz_1}\sim 500-1000$ GeV corresponds to the $\tz_1\tz_1\to A/H$
resonance annihilation region. This also should be accessible to LZ and DarkSide-20K but perhaps not to XENON1T.
The upper blue/gray region with $m_{\tz_1}\sim 500-1500$ GeV corresponds to the remnant HB/FP region
with a TeV-scale higgsino-like LSP. The LUX collaboration has excluded about half this parameter space
while LZ, XENONnT, DarkSide-20K and DEAP-50T should cover the remainder. 

The yellow band shows the locus of predictions for non-thermal wino dark matter~\cite{Baer:2010kd}
(as derived from our scans over the minimal anomaly-mediated SUSY breaking model or mAMSB)
using the IsaReS\cite{Baer:2003jb} subroutine of Isajet. 
A large chunk of parameter space has been ruled out by LUX. 
Since the neutralino-Higgs  coupling is proportional to
\be
X_{11}^h=-{1\over 2}(v_2^{(1)}\sin\alpha-v_1^{(1)}\cos\alpha)(gv_3^{(1)}-g'v_4^{(1)} )
\label{eq:X11}
\ee 
(where $v_2^{(1)}$ and $v_1^{(1)}$ are the two higgsino components of the $\chi_1$ and $v_3^{(1)}$ and $v_4^{(1)}$ 
are wino and bino components of $\chi_1$ in the notation of Eq. 8.117 of Ref.~\cite{Baer:2006rs} 
and $\alpha$ is the scalar Higgs mixing angle) we see the coupling is a product of
higgsino times gaugino components. When the $\chi_1$ becomes nearly pure wino ({\it e.g.} for light winos
but heavy scalars and large $\mu$), then the coupling in Eq. \ref{eq:X11} becomes very small and 
additional scattering contributions not included in IsaReS involving $W$-boson box diagrams
become important. These contributions have been evaluated in Ref's~\cite{Hisano:2010fy} and lead to a minimal
wino-proton scattering cross section which asymptotes around 
$\sigma^{SI}(\tz_1 p)\sim 2\times 10^{-47}$ cm$^2$ for $m(wino)\gtrsim 500$ GeV (this asymptotic limit
contains recently computed QCD corrections which increase the scattering cross section by $\sim 1.7$
compared to earlier results~\cite{Hisano:2015rsa}). 
This asymptotic limit (adapted from Ref.~\cite{Hisano:2015rsa})-- lying just above the neutrino floor-- 
implies that the NThW scenario, where wino-like neutralinos comprise the totality of
dark matter, will be completely explored by multi-ton noble liquid WIMP detectors.
For cases with lower $\mu$ values, then a mixed wino-higgsino LSP occurs and then the 
SI scattering rate is higher, and tends to be excluded.

The remaining model is RNS with a mainly higgsino-like LSP that constitutes only a fraction of the relic density.
The model prediction from the two-extra-parameter non-universal Higgs model (NUHM2) with $\Delta_{\rm EW}<15$
is shown by orange with $m_{\tz_1}\sim 100-250$ GeV (our scan). 
This region has only begun to be probed by recent LUX results
but should be fully explored by XENON1T, by LZ and by DarkSide-20K. 
The upper boundary of the region is determined by the LHC limit on 
gluino mass: $m_{\tg}\gtrsim 1.9$ TeV. The slightly more fine-tuned region with
$\Delta_{\rm EW}<30$ is shown in green. Assuming $\Omega_{\tz_1}^{\rm TP}h^2=\Omega_{\tz_1}h^2$, parts of this region may lie 
below XENON1T reach but should be accessible to XENONnT, LZ, DarkSide-20K and other ton-scale noble liquid detectors. In this 
region, a small fraction of dark matter ($\sim 10\%$) is comprised of higgsino-like neutralinos. In the Peccei-Quinn 
augmented SUSY scenario, non-thermal neutralino production from axino decays will augment neutralino abundance~\cite{Bae:2015jea}
hence the whole region might become accessible to XENON1T.

Finally, we also show the range of predictions in $\xi \sigma^{SI}(\tz_1, p)$ vs. $m_{\tz_1}$ space of the pMSSM 
analysis (region adapted from Fig. 5{\it a} of Ref.~\cite{Cahill-Rowley:2014boa}). 
We see the lower range starts around $m_{\tz_1}\sim 100$ GeV (a further small region exists around 
$m_{\tz_1}\sim m_Z/2$ and $m_h/2$ where bino resonance annihilation may occur) and encompasses all the theoretical model predictions, 
with $m_{\tz_1}$ ranging up to about 1.5 TeV. 
The latter limit is an artifact of the upper limits chosen for the scan over
pMSSM parameter space. 
Since the pMSSM includes all other models as subsets, it is perhaps not surprising that the model
encompasses all other predictions, and then some.

\section{Spin-dependent WIMP-nucleon scattering}
\label{sec:SD}
\begin{figure}[tbp]
\includegraphics[height=0.483\textheight]{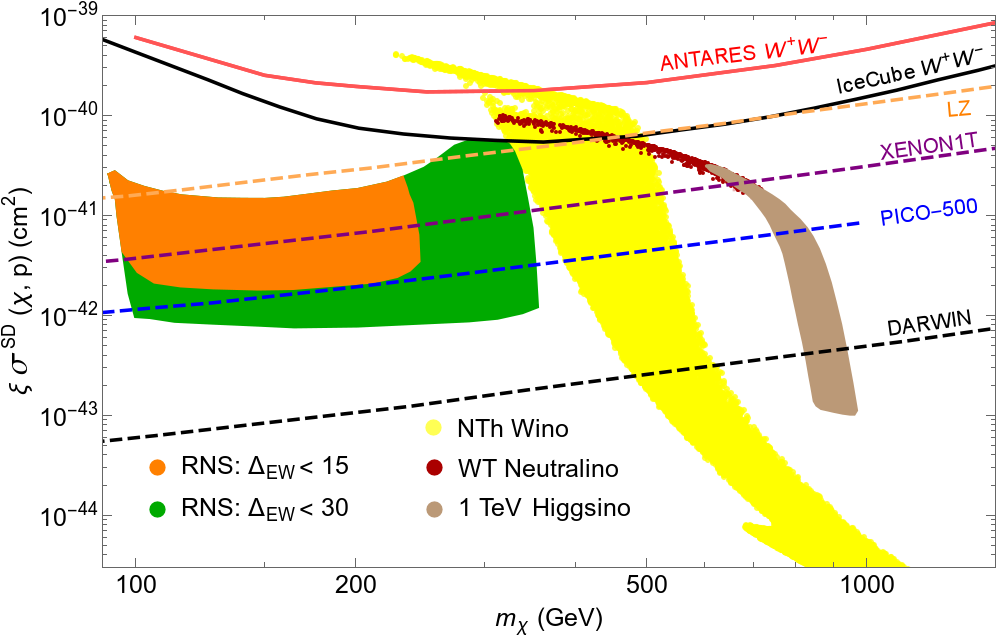}
\caption{Plot of rescaled spin-dependent WIMP detection rate 
$\xi \sigma^{SD}(\chi, p)$ 
versus $m_{\chi}$ from several published results versus current 
ANTARES and IceCube reach and projected (dashed) LZ, XENON1T, PICO-500 and DARWIN reaches. 
$\xi =1$ ({\it i.e.} it is assumed WIMPs comprise 
the totality of DM) for the experimental projections and for all models {\it except} RNS and pMSSM (not shown).
\label{fig:SD}}
\end{figure}
The spin-dependent WIMP-nucleon scattering cross section 
$\sigma^{SD}(\chi, p)$ vs. $m_{\chi}$ is shown in Fig.~\ref{fig:SD}.
These scattering reactions take place via $Z$ and squark exchange; again, since
squarks are expected heavy, the $Z$-exchange diagram should dominate. However, 
the $Z$-exchange coupling is proportional to (Eq. 8.101 of Ref.~\cite{Baer:2006rs})
\be
X_{11}^Z\sim {1\over 4}\sqrt{g^2+g^{\prime 2}}(v_1^{(1)2}+v_2^{(1)2})
\ee
which depends only on the higgsino components of $\tz_1$. Thus, models 
with a mainly higgsino-like LSP tend to yield large SD scattering cross sections. 
While a variety of underground experiments have developed bounds on $\sigma^{SD}$, the best recent
bounds come from the IceCube experiment which monitors WIMP annihilation into high energy neutrinos in 
the solar core. 
In most cases, the solar annihilation rate reaches equilibration with the solar WIMP capture
rate and the latter depends mainly on $\sigma^{SD}$. 
This is because the proton carries spin and there are plenty of protons within the sun 
to serve as targets for WIMP scattering and capture. 
The rate is relatively insensitive to $\sigma^{SI}$ since that rate requires enhancement 
by the number of nucleons in the nuclei.

The recent Antares search limit is shown by the red contour~\cite{Adrian-Martinez:2016gti} while
the recent IceCube limit is shown by the solid black contour~\cite{Aartsen:2016exj}. 
We see that IceCube rules out about half the WTN region and the upper portion of the yellow NThW region. 
For $m(wino)\simeq2$ TeV, $\sigma^{SD}$ extends down to $10^{-46}$ cm$^2$ which is well beyond any projected 
search limits. The IceCube limit barely touches the RNS model region because again RNS includes a depleted local abundance 
so that there simply may not be enough WIMPs around to become captured by the Sun.
We also show projected reaches of LZ~\cite{Akerib:2015cja}, XENON1T~\cite{Aalbers:2016jon}, 
Pico-500~\cite{ckrauss} and DARWIN~\cite{Aalbers:2016jon}.
The projected reach of LZ, shown by the orange dashed contour, will extend the reach for
$\sigma^{SD}$ into the lower mass WIMP range, which is already excluded by the recent SI LUX result, 
but may not reach much of the projected RNS parameter space.
No projections for $\sigma^{SD}$ vs. $m_{\tz_1}$ were found from the MC,  BF or FO collaborations.
However, we have generated the 1-TeV higgsino region using Isajet~\cite{Paige:2003mg} which is denoted 
with brown shading, assuming that $m_{1/2} \leq 5$ TeV. This region seems unlikely to be accessible to near future searches for SD scattering
but may be probed by Pico-500 and ultimately DARWIN.
The pMSSM predictions, shown in Fig. 5{\it b} of Ref. \cite{Cahill-Rowley:2014boa}, 
fill essentially the entire plane shown, so we do not show these here.

\section{Indirect detection of signals from WIMP-WIMP annihilation}
\label{sec:IDD}

In this section, we focus on some recent results from indirect detection of SUSY WIMP dark matter
via halo annihilation events $\tz_1\tz_1\to SM$ particles. There are a large assortment of final states
that can be searches for including, $\bar{p}$, $e^+$, $\bar{d}$, $\gamma$-line spectra and 
$\gamma$-continuum spectra. In addition, the expected signal rates are highly dependent on the assumed
dark matter density distribution. The portrait of theory vs. experiment is usually presented in the
thermally averaged cross section times velocity (in the limit as $v\to 0$) 
$\langle\sigma v\rangle$ vs. $m_{\chi}$ plane.
Here, we select out the $\chi_1\chi_1\to W^+W^-\to \gamma$ continuum limits since most of the SUSY models 
portrayed have this dominant annihilation channel (the exception being the stau and $A$ funnel annihilation 
regions from MC, BF and FO collaborations).

\begin{figure}[tbp]
\includegraphics[height=0.483\textheight]{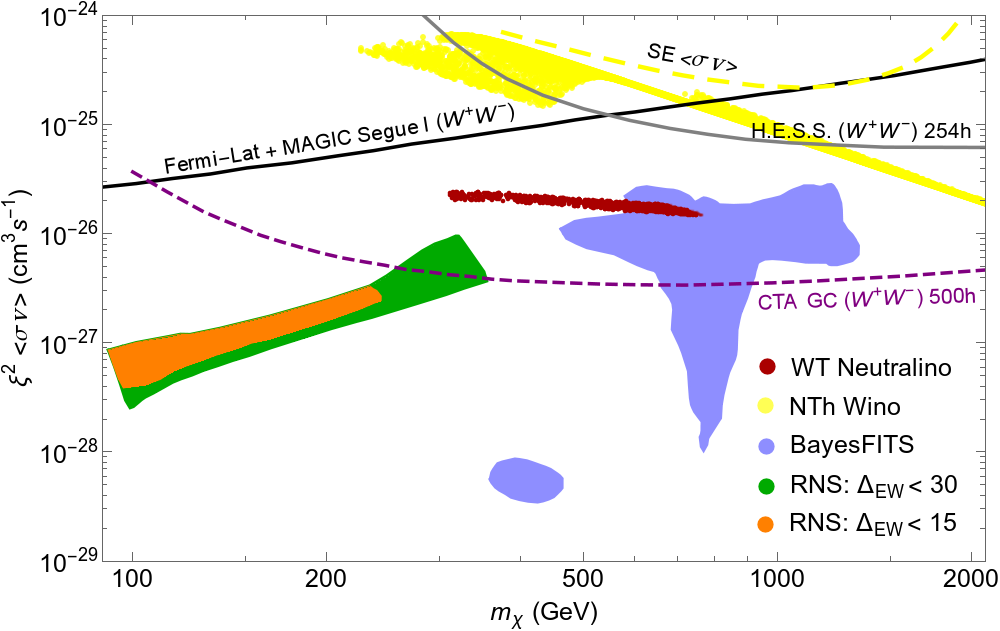}
\caption{Plot of rescaled thermally-averaged WIMP annihilation 
cross section times velocity $\xi^2 \langle \sigma v\rangle$ 
versus $m_{\chi}$ from several published results along with current
Fermi-LAT/MAGIC combined reach via $W^+W^-$ channel and projected (dashed) CTA reach. 
$\xi =1$ ({\it i.e.} it is assumed WIMPs comprise 
the totality of DM) for the experimental projections and for all models {\it except} RNS and pMSSM (not shown).
\label{fig:IDD}}
\end{figure}

The plane plot is shown in Fig.~\ref{fig:IDD}. We plot the recent combined Fermi-LAT+MAGIC limits found
from examining continuum gamma ray spectra from the dwarf spheroidal galaxy Segue I~\cite{Ahnen:2016qkx}.
In addition, we plot the updated 10 years/254 hours of HESS search for continuum gamma rays~\cite{::2016jja}.
We also show a projected gamma ray reach of the CTA collaboration assuming 500 hours of 
observation~\cite{Wood:2013taa}.

From the plot, we see that the maroon WTN,  while being excluded by SI direct detection searches, 
is still allowed in this IDD channel. 
The lower blue disjoint region  is stau co-annihilation (adapted from the BF collaboration Ref. \cite{Roszkowski:2014wqa}) 
while the upper blue region combines expectations from a 1 TeV higgsino LSP (upper half) 
with the $A/H$ resonance region (lower half). 
The 1 TeV higgsino-LSP should be testable by CTA~\cite{Roszkowski:2014wqa} 
even though the related gluino and squark masses are far beyond reach of LHC14. 

The RNS SUSY regions are suppressed by their $\xi^2$ factors in that the WIMPs 
may comprise only a fraction of the galactic dark matter abundance. 
Thus, their projected region of interest lies for the most part below even the CTA projected reach.
The pMSSM projections, given in Fig. 12 of Ref. \cite{Cahill-Rowley:2014boa}, fill essentially all of the parameter space shown.

Pertaining to NThW dark matter, we note that there have already been some claims in the
literature that these candidates are excluded by HESS and 
Fermi gamma-ray line searches~\cite{Cohen:2013ama,Fan:2013faa}. 
The reason NThWs are susceptible to such searches is that 1. the wino-wino$\to\gamma\gamma$ reaction proceeds
through a box diagram including wino-$W$ boson exchange and so is quite unsuppressed for wino-like WIMPs
and 2. Sommerfeld enhanced (SE) annihilation rates boost the annihilation cross section for higher mass winos.
These exclusion claims may be tempered by the more conservative analysis from Ref.~\cite{Hryczuk:2014hpa}
which maintains that winos are excluded for $m(wino)\lesssim 0.8$ TeV due to searches for $\bar{p}$s
and excluded between 1.8-3.5 TeV due to gamma-ray line searches. 
Thus, for Ref.~\cite{Hryczuk:2014hpa}, a window of viability remained open for 0.8 TeV $<m(wino)<1.8$ TeV. 

Our calculations from Isatools~\cite{Baer:2005bu} generate the expected $\langle\sigma v\rangle$ region from a scan over 
mAMSB models without SE as the yellow-shaded region.
We see that the continuum $\gamma$-ray search from the new combined Fermi/MAGIC/HESS results exclude
this scenario for $m(wino)\lesssim 1$ TeV. The dashed yellow line shows the expected SE value~\cite{Hryczuk:2011vi} 
of $\langle \sigma v\rangle$ which rises to a resonant maximum at $m(wino)\sim 2.4$ TeV after which it again falls. 
Including the Sommerfeld enhancement then seems to exclude wino dark matter via the continuum $\gamma$-ray searches
over values ranging up the $\sim 3$ TeV where the thermally-produced relic density then saturates 
the measured abundance (so no non-thermal enhancement is needed). 
Thus, NThW dark matter seems excluded by the new Fermi/MAGIC/HESS continuum $\gamma$-ray search results.
We do note here that mixed wino-axion dark matter still seems viable~\cite{Bae:2015rra}. 
In this case, the IDD rates are suppressed by $\xi^2$ factors which may range down to $\sim 10^{-4}$
which makes wino-wino halo annihilations rare just due to the paucity of winos compared to axions
in the galactic halo.

\section{Summary and conclusions}
\label{sec:conclude}

We summarize with a set of brief conclusions:
\begin{itemize}
\item The well-tempered neutralino is solidly excluded by recent XENON100, PandaX and LUX SI direct detection bounds.
\item The non-thermal wino which might comprise all dark matter was previously claimed to be excluded based
mainly on gamma ray line searches. It now seems also excluded by gamma ray continuum searches by Fermi-LAT/MAGIC 
combined with recent HESS results. 
It will also be probed completely via multi-ton noble liquid detectors via SI scattering.
The scenario of wino-like WIMP seems to survive if one postulates that the wino
{\it comprises only a fraction} of the dark matter~\cite{Fan:2013faa} with 
{\it e.g.} axions comprising the remainder~\cite{Bae:2015rra}.
\item Predictions from the CMSSM model have been strongly constrained by recent LUX SI DD limits although
broad sections of parameter space still survive. These all seem to have $m_{\chi}\gtrsim 350$ GeV.
Multi-ton noble liquid detectors will be needed to completely explore the allowed parameter space.
This model may already be considered not-so-pausible because the remaining parameter space gives rise to a 
$\mu$ parameter with $|\mu |\gg m_Z$: 
this can be interpreted as a poor prediction of $m_Z$ if fine-tuning had not been invoked.
\item The RNS models with small $\mu \lesssim 300$ GeV are natural and predict the existence of a higgsino-like LSP
that comprises only a fraction of the dark matter. The predicted parameter space, even accounting for a depleted
local abundance, is amenable to searches by ton-scale noble liquid detectors such as XENON, LZ, DarkSide, DEAP and
DARWIN. 
If naturalness in the QCD sector is eschewed so that the axion does {\it not} constitute the extra relic abundance, 
then non-thermal higgsino production must be invoked and the higgsinos
would comprise all dark matter with $\xi =1$. This case is already severely constrained by SI DD searches.
\item If XENON1T does not see a WIMP signal, 
the remaining parameter space for the CMSSM model (that saturates the measured dark matter abundance) 
predicts a heavy gluino mass $m_{\tilde{g}} \gtrsim 8$ TeV which is far above from expectations from a natural SUSY model. 
This lower limit on gluino mass applies for NUHM2 model with $\sim1$ TeV higgsino-like neutralino as well. RNS models with 
$m_{\tilde{g}}\lesssim4$ TeV will still 
survive since the SI detection rate is scaled down by the factor $\xi$. Furthermore, resonance annihilations such as 
$\chi_1\chi_1\to A/H$ would decrease the local WIMP abundance and push a substantial amount of the RNS region beyond XENON1T reach. Indeed 
for $m_{A/H} \simeq 2 m_{\tz_1}$, $\Omega_{\chi_1}^{\rm TP}h^2$ decreases by a factor of $\sim30$ but fortunately DarkSide-20K and 
DEAP-50T will eventually explore such regions. We expect additional contributions to the neutralino abundance 
from axino decays (which increases $\xi$); then a WIMP detection would be expected sooner.
\item If a WIMP signal is seen in the near future, then it will be highly useful to be able to distinguish its properties
based on mass and mixing. The case of ascertaining a WIMP mass $m_{\chi}\lesssim 350$ GeV (RNS) from the CMSSM case of 
$m_{\chi}\gtrsim 350$ GeV may be possible using mass measurement techniques and signals from different target 
materials~\cite{wimpmass}.
\item While many constrained SUSY models are indeed under seige from direct/indirect WIMP search experiments, the pMSSM--
with unconstrained soft parameters-- is typically less under seige. For instance, if the WIMP is nearly pure bino
with a diminished relic abundance such that $\Omega h^2(bino)\simeq 0.12$ 
(due perhaps to co-annihilation or resonance annihilation or entropy dilution) and all other sparticles are heavy and beyond
collider reach, then such scenarios yield very low direct/indirect detection rates. 
Such an unusual scenario might survive most or all search venues.
\item Detection of WIMPs or associated particles (in this case superpartners) at collider 
experiments will provide crucial information for distinguishing amongst the models considered here.
\end{itemize}

\section*{Acknowledgments}

We thank E. Aprile and M. Selvi for helpful suggestions and X. Tata for
reading the manuscript.
This work was supported in part by the US Department of Energy, Office of High
Energy Physics. HS would like to thank CETUP* (Center for Theoretical Underground Physics
and Related Areas) while this work was initiated. The computing for this project was 
performed at the OU Supercomputing Center for Education \& Research (OSCER) at the 
University of Oklahoma (OU).

%

%
\end{document}